\documentclass[preprint,pre,aps,amsfonts,amsmath,showpacs,superscriptaddress,nofootinbib]{revtex4}
\usepackage{bm}
\usepackage{graphicx}
\begin{document}
\title{Nonequilibrium Stationary State of a Truncated Stochastic Nonlinear Schr\"odinger Equation: I. Formulation and Mean Field Approximation}
\author{Philippe Mounaix}
\email{mounaix@cpht.polytechnique.fr}
\author{Pierre Collet}
\email{collet@cpht.polytechnique.fr}
\affiliation{Centre de Physique Th\'eorique, UMR 7644 du CNRS, Ecole
Polytechnique, 91128 Palaiseau Cedex, France.}
\author{Joel L. Lebowitz}
\email{lebowitz@math.rutgers.edu}
\affiliation{Departments of Mathematics and Physics, Rutgers, The State
University of New Jersey, Piscataway, New Jersey 08854-8019.}
\date{\today}
\begin{abstract}
We investigate the stationary state of a model system evolving according to a modified focusing truncated nonlinear Schr\"odinger equation (NLSE) used to describe the envelope of Langmuir waves in a plasma. We restrict the system to have a finite number of normal modes each of which is in contact with a Langevin heat bath at temperature $T$. Arbitrarily large realizations of the field are prevented by restricting each mode to a maximum amplitude. We consider a simple modeling of wave-breaking in which each mode is set equal to zero when it reaches its maximum amplitude. Without wave-breaking the stationary state is given by a Gibbs measure. With wave-breaking the system attains a nonequilibrium stationary state which is the unique invariant measure of the time evolution. A mean field analysis shows that the system exhibits a transition from a regime of low field values at small $\vert\lambda\vert$, to a regime of higher field values at large $\vert\lambda\vert$, where $\lambda<0$ specifies the strength of the nonlinearity in the focusing case. Field values at large $\vert\lambda\vert$ are significantly smaller with wave-breaking than without wave-breaking.
\end{abstract}
\maketitle
%
%%%%%%%%%%%%%%%%%%%%
%
\section{Introduction}\label{sec1}
The statistical mechanics of continuum fields is a problem of both intrinsic and applied interest. It is often at the level of the coarse-grained continuum description provided by the collective modes of the system, rather than at the underlying level of the atomic degrees of freedom, where certain types of physical phenomena are most clearly manifest. Thus, turbulence in fluids is best described by statistical properties of hydrodynamic fields evolving according to the Navier-Stokes equations..
\paragraph*{}Another example, relevant for plasmas, is the case of a complex field $\phi$ the dynamics of which is governed by the nonlinear Schr\"odinger equation:
\begin{equation}\label{eq1.1}
\left\lbrace
\begin{array}{l}
i\partial_t\phi+\Delta\phi=
\lambda\vert\phi\vert^{p-2}\phi ,\\[1ex]
t\ge 0,\ x\in\Lambda\subset {\mathbb R}^D,\ {\rm and}\ \phi(x,t=0)=\phi_0(x) ,
\end{array}
\right.
\end{equation}
where $\lambda\in {\mathbb R}$ is a parameter specifying the strength of the nonlinearity, $p >2$ is an integer, and $\Lambda$ is a compact domain in ${\mathbb R}^D$ which we shall generally take to be a torus. Equation\ (\ref{eq1.1}) with $\lambda <0$ (the focusing case) and $p=4$ has been used to describe some of the physics associated with the slowly varying envelope of Langmuir waves in a plasma, such as wave collapse in the subsonic regime\ \cite{MG}, as well as other physical phenomena\ \cite{NLS}. For $\lambda <0$ the Hamiltonian generating the dynamics,
\begin{equation}\label{eq1.2}
\hat{{\mathcal H}}(\phi)=\int_{\Lambda}\left(
\frac{1}{2}\vert\nabla\phi\vert^2 +\frac{\lambda}{p}\vert\phi\vert^p\right)\, d^Dx,
\end{equation}
is not bounded below. Consequently there does not exist a Gibbs measure for the fields evolving according to\ (\ref{eq1.1}). One way of overcoming this problem is by putting bounds on the conserved ``mass",
\begin{equation}\label{eq1.3}
\hat{{\mathcal N}}(\phi)=\frac{1}{4}\int_{\Lambda}\vert\phi\vert^2\, d^Dx.
\end{equation}
This is done in\ \cite{LRS} for $D=1$, see also\ \cite{Bou1}, where a well-defined measure is constructed for parameter ranges where there is no wave collapse\ \cite{MG}. This approach is difficult to extend to $D>1$ where one is faced with ultraviolet problems, see\ \cite{Bou2} for the defocusing case, and is not suitable to cases where wave collapse can occur.
\paragraph*{}In this paper we investigate the properties of a field $\phi$ evolving according to a truncated version of\ (\ref{eq1.1}) with added stochastic terms as well as an upper bound on the amplitude of the Fourier components of $\phi$. The stochastic Langevin forces and the focusing nonlinearity want to drive the field to arbitrarily large values, which is prevented by the bound on the amplitude. We also consider a mechanism which resets the amplitude of a Fourier mode to zero when it reaches its maximum. This mechanism imitates, in a simplified way, the wave-breaking of Langmuir waves in a plasma\ \cite{WK}\ \cite{TC}, which we now describe.
\paragraph*{}The wave-breaking of a Langmuir wave is the violent dissipation of its energy when its amplitude is high enough for the bulk of the thermal electrons to get trapped in the electrostatic potential of the wave\ \cite{WK}\ \cite{TC}. The larger the wave number $k$ the smaller the maximum amplitude the wave can reach. For $k\gtrsim\lambda_D^{-1}$, where $\lambda_D$ is the Debye screening length, wave-breaking turns into strong Landau damping so the amplitude does not grow. Wave-breaking introduces both a natural ultraviolet cut-off, $k_{max}\simeq\lambda_D^{-1}$, and a natural upper bound for the amplitude of each Fourier mode of $\phi$. Our modeling of wave-breaking is a discontinuous dissipative process whereby each Fourier component of $\phi$ is set equal to zero when it reaches a given amplitude. In the absence of wave-breaking the stationary measure is Gibbsean at the temperature imposed by the Langevin heat bath. With wave-breaking the stationary state is not an equilibrium one. We show however that it exists and is unique.
\paragraph*{}The outline of the paper is as follows. In Section\ \ref{sec2} we specify the stochastic  dynamics of our model system. This dynamics acting on a finite number of bounded degrees of freedom (the Fourier modes) ensure the existence of a unique stationary state which is approached exponentially fast. This is proven in Section\ \ref{sec3}. A mean field theory with and without wave-breaking is developed in Section\ \ref{sec4}. Consequences of an infrared divergence of the maximum amplitude in the mean field theory with wave-breaking are briefly addressed in Section\ \ref{sec4d}. Finally, the relevance of the mean field theory and some possible improvements of our wave-breaking prescription are discussed in Section\ \ref{sec5}.
%
%%%%%%%%%%%%%%%%%%%%
%
\section{Model and definitions}\label{sec2}
Let $\Lambda$ be a $D$-dimensional torus of length $L$ and volume $V=L^D$. From now until Section\ \ref{sec4c}, we will take $L=1$ without loss of generality. Let $\eta_c>0$ be an ultraviolet cut-off [in plasma physics, $\eta_c\simeq(2\pi\lambda_D)^{-1}$], i.e. consider a field $\phi$ of the form
\begin{equation}\label{eq2.1}
\phi(x,t)=\sum_{\vert\vert n\vert\vert <\eta_c}a_n(t)\exp(ik_n\cdot x),
\end{equation}
with $k_n=2\pi n$, $n\in {\mathbb Z}^D$, and $\vert\vert n\vert\vert =(\sum_{i=1}^Dn_i^2)^{1/2}$. The truncated Hamiltonian and mass are given respectively by
\begin{equation}\label{eq2.2}
{\mathcal H}(a)\equiv {\mathcal H}(\phi)=
\frac{1}{2}\sum_{\vert\vert n\vert\vert <\eta_c} k_n^2\vert a_n\vert^2
+\frac{\lambda}{p}\int_{\Lambda}\left\vert
\sum_{\vert\vert n\vert\vert <\eta_c} a_n\exp(ik_n\cdot x)
\right\vert^p\, d^Dx,
\end{equation}
\begin{equation}\label{eq2.2b}
{\mathcal N}(a)\equiv {\mathcal N}(\phi)=
\frac{1}{4}\sum_{\vert\vert n\vert\vert <\eta_c}\vert a_n\vert^2.
\end{equation}
Note that ${\mathcal H}(a)$ and ${\mathcal N}(a)$ are different from their untruncated counterparts $\hat{{\mathcal H}}(a)$ and $\hat{{\mathcal N}}(a)$ defined in\ (\ref{eq1.2}) and\ (\ref{eq1.3}). In the absence of a bound on $\vert a_n(t)\vert$, the time evolution of the $a_n$ is assumed to be given by the truncated nonlinear Schr\"odinger equation\ (\ref{eq1.1}) combined with stochastic Langevin forces:
\begin{equation}\label{eq2.3}
\left\lbrace
\begin{array}{l}
\dot{a}_{nR}=\partial{\mathcal H}(a)/\partial a_{nI}-
\nu_n\partial\lbrack\mu{\mathcal N}(a)+{\mathcal H}(a)\rbrack/\partial a_{nR}+
\sqrt{2\nu_n\beta^{-1}}L_{nR},\\
\dot{a}_{nI}=-\partial{\mathcal H}(a)/\partial a_{nR}-
\nu_n\partial\lbrack\mu{\mathcal N}(a)+{\mathcal H}(a)\rbrack/\partial a_{nI}+
\sqrt{2\nu_n\beta^{-1}}L_{nI},
\end{array}
\right.
\end{equation}
where ${a}_{nR}={\rm Re}(a_n)$,  ${a}_{nI}={\rm Im}(a_n)$, $\nu_n$ is the linear damping of $a_n$, $\beta =(\kappa_BT)^{-1}$, $T$ is the temperature, $\mu$ is the chemical potential, and the $L_n$ are independent Gaussian white noises with $\langle L_{nR}(t)L_{nI}(t')\rangle =0$ and $\langle L_{nR}(t)L_{nR}(t')\rangle =\langle L_{nI}(t)L_{nI}(t')\rangle =\delta(t-t')$.
\paragraph*{}The motivations for\ (\ref{eq2.3}) are twofold : first, in the limit of no contact with the thermostat (i.e. $\nu_n=0$ for all $n$), it reduces to a truncated version of\ (\ref{eq1.1}) with $\phi$ of the form\ (\ref{eq2.1}) ; secondly, in the well defined defocusing case, $\lambda>0$, with an ultraviolet cut-off on $n$, the stationary probability distribution of the $a_n$ is the Gibbs distribution $P_{eq}(a)\propto {\rm e}^{-\beta\lbrack{\mathcal H}(a)+\mu{\mathcal N}(a)\rbrack}$, where $\beta$ and $\mu$ are specified by the Langevin terms\ \cite{Brown}. These can be thought of as representing the interaction between the waves and the particle degrees of freedom, not included in\ (\ref{eq2.2}). Even with this truncation the Hamiltonian ${\mathcal H}(a)$ in\ (\ref{eq2.2}) is still not bounded below for the case $\lambda <0$, the focusing case. Hence the dynamics in\ (\ref{eq2.3}) will not lead to a stationary state: the reservoirs would just keep extracting energy from the waves, causing  ${\mathcal H}(a)$ to go towards $-\infty$. To prevent this we let $\alpha_n$ be a monotone decreasing function of $\vert\vert n\vert\vert$ such that $\alpha_n=0$ for all $n$ with $\vert\vert n\vert\vert\ge\eta_c$ and $\alpha_n>0$ otherwise. Now, we assume that, as soon as $\vert a_n\vert =\alpha_n$, it is either reflected, i.e. $\dot{a}_n$ is set equal to $-\dot{a}_n$, or instantaneously set equal to $a_n=0$ (``wave-breaking" prescription) from where it resumes its time evolution according to\ (\ref{eq2.3}). These prescriptions are equivalent to imposing either a reflecting or an absorbing boundary condition at $\vert a_n\vert =\alpha_n$ on the Fokker-Planck equation associated with\ (\ref{eq2.3}). In the latter case we compensate exactly for the absorption with sources at $a_n=0$.
\paragraph*{}The probability distribution $P(a,t)$ of the above system satisfies the modified Fokker-Planck equation
\begin{equation}\label{eq2.4}
\partial_t P(a,t)-\hat{L}P(a,t)=
\varepsilon\sum_{\vert\vert n\vert\vert <\eta_c} \sigma_n\left(\lbrace a_{m\ne n}\rbrace ,t\right)\, 
\delta(a_{nR})\delta(a_{nI}),
\end{equation}
with $\varepsilon =0$ in the reflecting case and $\varepsilon =1$ in the wave-breaking case. The Fokker-Planck operator $\hat{L}$ is given by
\begin{equation}\label{eq2.5}
\hat{L}P(a,t)=-\sum_{\vert\vert n\vert\vert <\eta_c}\nabla_n\cdot J_n(a,t),
\end{equation}
with the probability current
\begin{eqnarray}\label{eq2.6}
J_n(a,t)&=&\left\lbrace\nabla_n\times{\mathcal H}(a)
-\nu_n\nabla_n\lbrack {\mathcal H}(a)+\mu{\mathcal N}(a)\rbrack
\right\rbrace P(a,t) \nonumber \\
&-&\nu_n\beta^{-1}\nabla_nP(a,t).
\end{eqnarray}
The sources $\sigma_n\left(\lbrace a_{m\ne n}\rbrace ,t\right)$ are given by
\begin{equation}\label{eq2.7}
\sigma_n\left(\lbrace a_{m\ne n}\rbrace ,t\right)=
-\nu_n\beta^{-1}\int_{\vert a_n\vert =\alpha_n}
e_n\cdot\nabla_nP(a,t)\, ds.
\end{equation}
Here we have written $\nabla_n=(\partial/\partial a_{nR},\partial/\partial a_{nI})$, $\nabla_n\times{\mathcal H}(a)=\lbrack\partial{\mathcal H}(a)/\partial a_{nI},-\partial{\mathcal H}(a)/\partial a_{nR}\rbrack$, $e_n=(a_{nR}/\vert a_n\vert,a_{nI}/\vert a_n\vert)$ and $ds=\sqrt{da_{nR}^2+da_{nI}^2}$. The boundary conditions on the solution to\ (\ref{eq2.4}) are $e_n\cdot J_n(a,t)\vert_{\vert a_n\vert =\alpha_n,\, \vert a_{m\ne n}\vert\le\alpha_m}=0$ for all $t\ge 0$ in the reflecting case, and $P(a,t)\vert_{\vert a_n\vert =\alpha_n,\, \vert a_{m\ne n}\vert\le\alpha_m}=0$ for all $t\ge 0$ in the wave-breaking case. We also have $P(a,t)\vert_{\vert a_n\vert >\alpha_n}=0$ for all $n$ and all $t\ge 0$ in both cases.
%
%%%%%%%%%%%%%%%%%%%%
%
\section{Exponential approach to a unique stationary solution}\label{sec3}
\begin{figure}[htbp]
\begin{center}
\includegraphics [width=8cm] {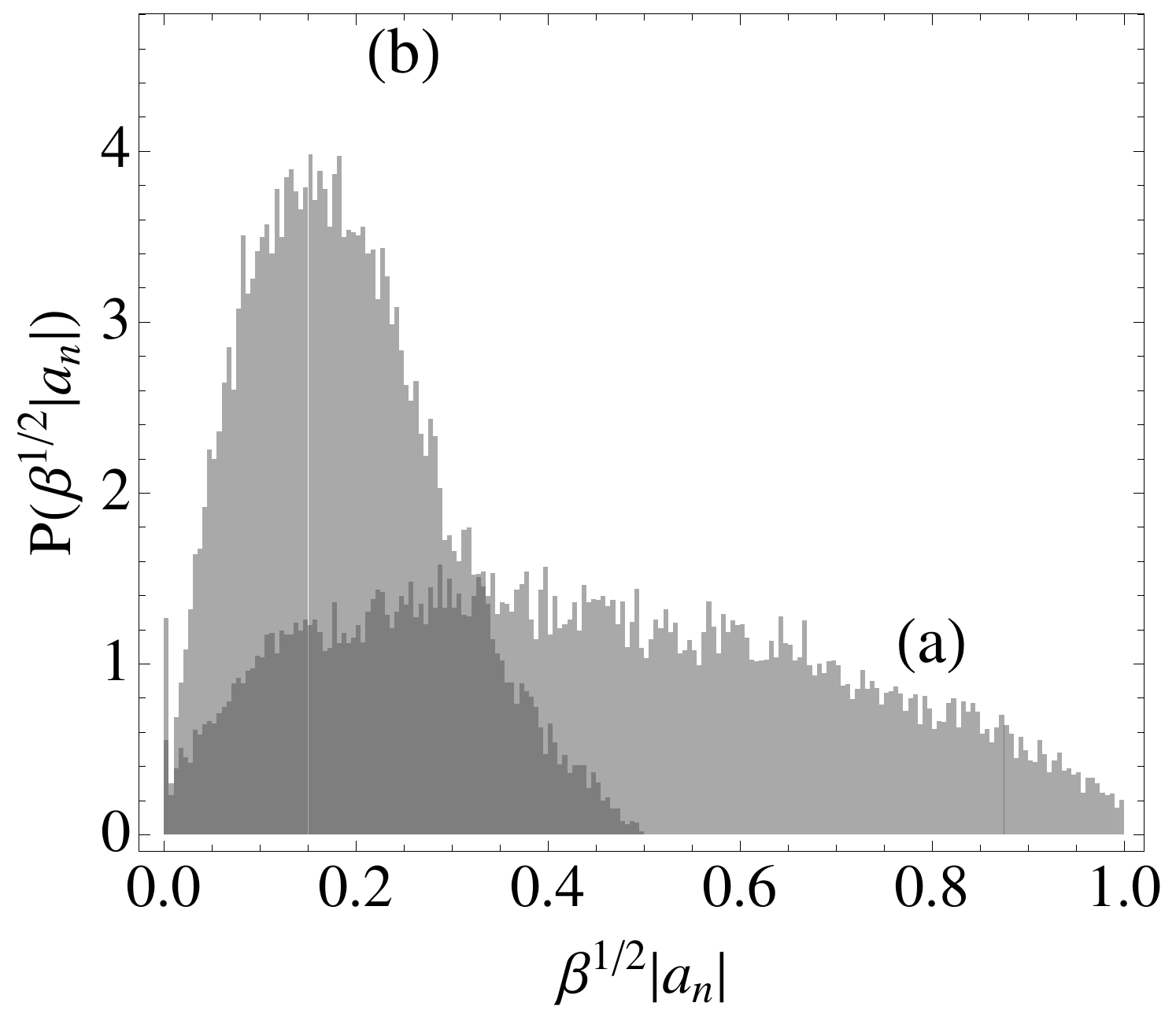}
\caption{\textsl{\small Stationary distribution of (a) $\beta^{1/2}\vert a_0\vert$ and (b) $\beta^{1/2}\vert a_1\vert$ obtained from a numerical solution of\ (\ref{eq2.3}) for a three-mode system ($n=0,\, \pm 1$) with $D=1$, $p=4$, $\lambda\beta^{-1}=-5$, $\nu_n=10$, $\beta^{1/2}\alpha_0=1$, and $\beta^{1/2}\alpha_{\pm 1}=1/2$. The distribution for $\beta^{1/2}\vert a_{-1}\vert$ is similar to (b).}}
\label{newfig0a}
\end{center}
\end{figure}

\begin{figure}[htbp]
\begin{center}
\includegraphics [width=8cm] {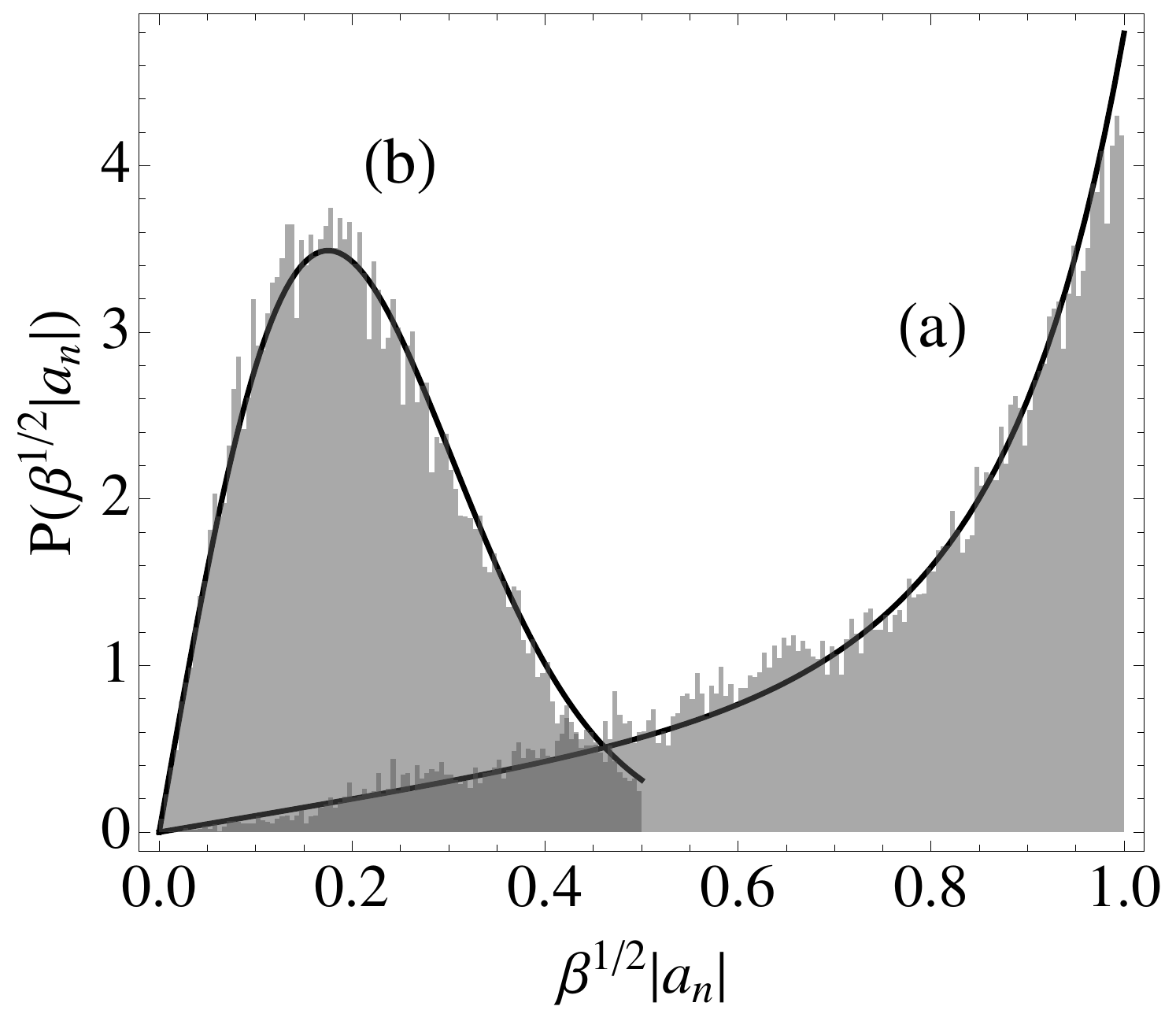}
\caption{\textsl{\small Stationary distribution of $\beta^{1/2}\vert a_n\vert$ obtained from a numerical solution of\ (\ref{eq2.3}) with reflecting boundary conditions, and analytical results obtained from the corresponding Gibbs distribution (plain lines). Parameters and notation are the same as in Fig.\ \ref{newfig0a}.}}
\label{newfig0b}
\end{center}
\end{figure}

In this section we show that the system with time evolution described by\ (\ref{eq2.4}) has a unique stationary state, $P_{st}(a)$, and that every initial probability distribution tends to $P_{st}(a)$ exponentially fast. For the reflecting case $P_{st}(a)$ is the equilibrium Gibbs state, $P_{st}(a)\propto {\rm e}^{-\beta\lbrack{\mathcal H}(a)+\mu{\mathcal N}(a)\rbrack}$ for $\vert a_n\vert <\alpha_n$, while for the wave-breaking prescription $P_{st}(a)$ will be a nonequilibrium stationary state which is unknown in general. Here we give the proof for the wave-breaking prescription only as the equilibrium case is well known\ \cite{Brown}.
\paragraph*{}The Langevin equations\ (\ref{eq2.3}) with the wave-breaking prescription defines a Markov process on the compact domain $ \Gamma\equiv {\rm supp}\prod_{\vert\vert n\vert\vert <\eta_c}\bm{1}_{\lbrace \vert a_n\vert <\alpha_n\rbrace}$. Let $P(a,t;b)$, with $a,b\in\Gamma$, denote the solution to\ (\ref{eq2.4}) with $P(a,0;b)=\delta (a-b)$ (with obvious notation), and write $P(A,t;b)=\int_{a\in A}P(a,t;b)\, d^Na$. D{\oe}blin's condition\ \cite{Var} says that if
\begin{equation}\label{eq3.1}
\rho(t)\equiv \sup_{b,b^\prime \in\Gamma ,\, A\subset\Gamma}
\vert P(A,t;b)-P(A,t;b^\prime)\vert <1,
\end{equation}
for some $t>0$, then there is a unique invariant measure with exponentially fast convergence to it, starting from any point in $\Gamma$. This follows from the fact that, using the Chapman-Kolmogorov equation for Markov processes, we have $\rho(t+s)\le\rho(t)\rho(s)$\ \cite{Var}. This implies in turn that if $\rho(t)$ satisfies\ (\ref{eq3.1}) for some $t>0$ then $\rho(t)\rightarrow 0$ exponentially as $t\rightarrow +\infty$. In order for\ (\ref{eq3.1}) to be violated it is necessary that ther exists a set $A$ and a configuration $b$ for which $P(A,t;b)=0$ for all $t>0$ and of course $P(A^c,t;b)=1$. Hence to prove\ (\ref{eq3.1}) it is clearly sufficient to show that
\begin{equation}\label{eq3.2}
\inf_{b\in\Gamma}P(a,t;b)\ge c>0
\end{equation}
for every $b\in\Gamma$ and some $t=t_0>0$, and all $a$ in some compact set $K\subset\Gamma$ of strictly positive Lebesgue measure. When this is true then clearly $c\vert A\cap K\vert\le P(A,t;b)\le 1-c\vert A^c\cap K\vert$ and $\rho(t)\le 1-c\vert K\vert <1$.
\paragraph*{}To prove\ (\ref{eq3.2}) for our system we simply note that for $a$ and $b$ away from $\partial\Gamma$, the boundary of $\Gamma$, i.e. $\vert a_n\vert <\alpha_n$, we always have $P(a,1;b)\ge P_{diff}(a,1;b)$ where $P_{diff}(a,t;b)$ is the transition probability density for the evolution described by\ (\ref{eq2.4}) without sources (i.e. $\sigma_n=0$ for all $n$, and absorbing boundaries at $\vert a_n\vert =\alpha_n$). This evolution would not conserve mass but $ P_{diff}(a,1;b)\ge c>0$ for $a$, $b$ away from $\partial\Gamma$. Letting now $b$ be arbitrary close to $\partial\Gamma$ one has $P(a,1;b)\rightarrow P(a,1;b_{pr})$ where $b_{pr}$ is the projection of $b$ according to the wave-breaking prescription, which is away from $\partial\Gamma$, and the proof is complete. (We are indebted to S. Varadhan for this streamlined argument).
\paragraph*{}As an illustration, Figure\ \ref{newfig0a} shows the stationary distribution of $\beta^{1/2}\vert a_n\vert$ obtained from a numerical solution of\ (\ref{eq2.3}) for a three-mode system ($n=0,\, \pm 1$) with $D=1$, $p=4$, $\lambda\beta^{-1}=-5$, $\nu_n=10$, $\beta^{1/2}\alpha_0=1$, and $\beta^{1/2}\alpha_{\pm 1}=1/2$. For each mode the stationary distribution is computed from the relative frequency of $\beta^{1/2}\vert a_n\vert\in\lbrack 5m\, 10^{-3},\, 5(m+1)\, 10^{-3}\lbrack$, with $m$ an integer, over a long enough period of time. The time at which we start the sampling is chosen large enough so that the results do not depend on its value. Figure\ \ref{newfig0b} is the counterpart of Figure\ \ref{newfig0a} for reflecting boundary conditions. Analytical results obtained from the corresponding Gibbs distribution are also shown (plain lines).
\bigskip
In this particular case the effects of the wave-breaking prescription are most clearly seen in the distribution of $\vert a_0\vert$. For reflecting boundary conditions (Fig.\ \ref{newfig0b}), the distribution of $\vert a_0\vert$ is maximum at the boundary, which means that this mode tends to stay in the vicinity of its maximum amplitude for a relatively long period of time. On the other hand (Fig.\ \ref{newfig0a}), the wave-breaking prescription prevents the mode from staying near its maximum amplitude very long. Instead, it is quickly reset to $\vert a_0\vert =0$, which results in a dramatic reduction of the distribution near the boundary and the appearance of a narrow peak at $\vert a_0\vert =0$.
\bigskip
%
%
%
%%%%%%%%%%%%%%%%%%%%
%
\section{Mean field theory ($\bm{p=4}$)}\label{sec4}
We turn now to a mean field approximation similar to that considered by various authors in various contexts (see e.g.\ \cite{LRS2}). It corresponds to replacing $\frac{1}{p}\vert\phi(x)\vert^p$ in\ (\ref{eq1.2}) with $\frac{1}{2}W(x)^{p/2-1}\vert\phi(x)\vert^2$ where $W(x)$ is to be determined self-consistently in a way to be discussed below. In the following, we will take $p=4$, assume that $W(x)\equiv W$ is independent of $x$, and consider both the reflecting ($\varepsilon =0$) and wave-breaking ($\varepsilon =1$) cases.
%
%%%%%%%%%%%%%%%%%%%%
%
\subsection{General results}\label{sec4a}
Our mean field theory is defined by replacing ${\mathcal H}(\phi)$ with
\begin{equation}\label{eq4.2}
{\mathcal H}_{MF}(\phi)=\int_{\Lambda}\left(
\frac{1}{2}\vert\nabla\phi\vert^2
+\frac{\lambda}{2}W\vert\phi\vert^2\right)\, d^Dx.
\end{equation}
In this approximation, the Fourier components of $\phi$ are decoupled and evolve according to\ (\ref{eq2.3}) in which ${\mathcal H}(a)$ is replaced with
\begin{equation}\label{eq4.3}
{\mathcal H_{MF}}(a)\equiv {\mathcal H}_{MF}(\phi)=\frac{1}{2}\sum_{\vert\vert n\vert\vert <\eta_c}
\left(k_n^2+\lambda W\right)\vert a_n\vert^2.
\end{equation}
According to\ (\ref{eq2.4}), the mean field stationary measure of $a_n$ is the solution to
\begin{eqnarray}\label{eq4.5}
0&=&-\frac{\partial}{\partial a_{nR}}\left\lbrack\left(
\frac{\partial {\mathcal H}_{MF}(a)}{\partial a_{nI}}-
\nu_n\frac{\partial {\mathcal H}_{MF}(a)+\mu {\mathcal N}(a)}{\partial a_{nR}}\right)
P(a_n)\right\rbrack \nonumber \\
&+&\frac{\partial}{\partial a_{nI}}\left\lbrack\left(
\frac{\partial {\mathcal H}_{MF}(a)}{\partial a_{nR}}+
\nu_n\frac{\partial {\mathcal H}_{MF}(a)+\mu {\mathcal N}(a)}{\partial a_{nI}}\right)
P(a_n)\right\rbrack \\
&+&\nu_n\beta^{-1}\left(\frac{\partial^2}{\partial a_{nR}^2}+
\frac{\partial^2}{\partial a_{nI}^2}\right)P(a_n) +\varepsilon\sigma_n\delta(a_{nR})\delta(a_{nI}), \nonumber
\end{eqnarray}
with the boundary conditions
\begin{equation}\label{eq4.6}
\begin{array}{ll}
e_n\cdot J_n(a_n)\vert_{\vert a_n\vert =\alpha_n}=0,&\ {\rm for}\ \varepsilon =0,\\
P(a_n)\vert_{\vert a_n\vert =\alpha_n}=0,&\ {\rm for}\ \varepsilon =1,
\end{array}
\end{equation}
and where
\begin{equation}\label{eq4.7}
\sigma_n=
-\nu_n\beta^{-1}\int_{\vert a_n\vert =\alpha_n}
e_n\cdot\nabla_nP(a_n)\, ds.
\end{equation}
Let $P_{MF}(a_n)$ denote this stationary measure. One has,
\begin{eqnarray}\label{eq4.8}
P_{MF}(a_n)&=&\frac{\bm{1}_{\vert a_n\vert\le\alpha_n}}{Z_n(W)}\exp\left\lbrack -\frac{\beta}{2}
\left(k_n^2+\lambda W+\frac{\mu}{2}\right)\vert a_n\vert^2\right\rbrack \nonumber \\
&\times&\int_{\varepsilon\vert a_n\vert /\alpha_n}^{1}
\frac{1}{r^\varepsilon}\exp\left\lbrack\frac{\varepsilon\beta}{2}
\left(k_n^2+\lambda W+\frac{\mu}{2}\right) \alpha_n^2 r^2\right\rbrack\, dr,
\end{eqnarray}
where $Z_n(W)$ is a normalization constant such that $\int_{\vert a_n\vert\le\alpha_n}P_{MF}(a_n)\, da_{nR}da_{nI}=1$, and the integral over $r$ is equal to $1$ when $\varepsilon =0$. Before writing the expression of $Z_n(W)$ it is convenient to introduce the function
\begin{equation}\label{eq4.9}
h_n(x)=\frac{\beta}{2}\left(k_n^2+\lambda x+\frac{\mu}{2}\right).
\end{equation}
After some algebra one finds that $Z_n(W)$ is given by
\begin{equation}\label{eq4.10}
Z_n(W)=\frac{\pi}{h_n(W)}\left\lbrack 1-\exp(-\alpha_n^2h_n(W))\right\rbrack ,
\end{equation}
in the reflecting case ($\varepsilon =0$), and by
\begin{equation}\label{eq4.11}
Z_n(W)=\left\lbrace
\begin{array}{ll}
\frac{\pi}{2h_n(W)}\left\lbrack
{\rm Ei}(\alpha_n^2h_n(W))-\ln(\alpha_n^2h_n(W))-\gamma
\right\rbrack &{\rm for}\ \ h_n(W)\ge 0, \\
\frac{\pi}{2\vert h_n(W)\vert}\left\lbrack
{\rm E}_1(\alpha_n^2\vert h_n(W)\vert)+\ln(\alpha_n^2\vert h_n(W)\vert)+\gamma
\right\rbrack &{\rm for}\ \ h_n(W)\le 0,
\end{array}
\right.
\end{equation}
where $\gamma$ is the Euler constant, in the wave-breaking case ($\varepsilon =1$).
%
%
%%%%%%%%%%%%%%%%%%%%
%
\subsection{Self-consistency and nonuniqueness}\label{sec4b}
There are various self-consistent prescriptions for finding $W$ and it is a matter of judgement which gives the best approximation. Two ``natural" ways to impose self-consistency are: (i) require that the average energies agree, which yields
\begin{equation*}
\frac{1}{2}W^{p/2-1}\langle\vert\phi\vert^2\rangle_W =\frac{1}{p}\langle\vert\phi\vert^p\rangle_W,
\end{equation*}
where $\langle\cdot\rangle_W$ denotes the average w.r.t. the stationary mean field measure; and (ii) require that in the dynamics specified by\ (\ref{eq1.1}),
\begin{equation*}
W^{p/2-1}\phi =\langle\vert\phi\vert^{p-2}\rangle_W \phi.
\end{equation*}
In the equilibrium case corresponding to $\varepsilon =0$, a third ``natural" prescription is: (iii) require that $W$ minimize the free energy\ (\ref{eq4.15}).
\paragraph*{}For the case $p=4$, $\lambda >0$, $\alpha_{n,\, \| n\| <\eta_c}=+\infty$, and a Gaussian stationary measure, these prescriptions give the self-consistency equation $W=q\langle\vert\phi\vert^2\rangle_W$ with $q=1$ for (i) and (ii), and $q=2$ for (iii).
\paragraph*{}We shall use the same self-consistency equation in our model also for $\lambda <0$ and finite $\alpha_{n,\, \| n\| <\eta_c}$, giving $W=q\sum_{\vert\vert n\vert\vert <\eta_c}\langle\vert a_n\vert^2\rangle_W$ with $\left\langle\vert a_n\vert^2\right\rangle_W =\int_{\vert a_n\vert\le\alpha_n}\vert a_n\vert^2P_{MF}(a_n)\, da_{nR}da_{nI}$. We shall leave $q$ as a free parameter from now on as the results depend on the choice of $q$ only via a rescaling of $\lambda$. Using\ (\ref{eq4.8}),\ (\ref{eq4.10}), and\ (\ref{eq4.11}), this yields the self-consistency equation
\begin{equation}\label{eq4.14}
W=q\sum_{\vert\vert n\vert\vert <\eta_c}
\frac{1}{h_n(W)}\left\lbrack 1-\frac{\pi\alpha_n^2}{2^\varepsilon Z_n(W)}\exp((\varepsilon -1)\alpha_n^2h_n(W))\right\rbrack .
\end{equation}
\paragraph*{}We now consider the case where\ (\ref{eq4.14}) has more than one solution. We then need to determine which one is to be picked. In the reflecting case this can be done by choosing the solution with the smallest free energy
\begin{equation}\label{eq4.15}
\frac{\lambda}{4}\left\langle\left\vert\sum_{\vert\vert n\vert\vert <\eta_c}a_n\right\vert^4\right\rangle_W
+\frac{1}{\beta}\int P_{MF}(a)\ln P_{MF}(a)\, da,
\end{equation}
which, up to a term independent of $W$, is equal to
\begin{equation}\label{eq4.16}
\lambda\left(\frac{1}{4}\left\langle\left\vert\sum_{\vert\vert n\vert\vert <\eta_c}a_n\right\vert^4\right\rangle_W
-\frac{1}{2}W\sum_{\vert\vert n\vert\vert <\eta_c}\left\langle\vert a_n\vert^2\right\rangle_W\right)
-\frac{1}{\beta}\sum_{\vert\vert n\vert\vert <\eta_c}\ln Z_n(W).
\end{equation}
As mentioned above, one could alternatively define $W$, in the reflecting case, as the value minimizing\ (\ref{eq4.16}). It will be shown later that, in the large system limit, this definition of $W$ does not give significantly different results from our definition with $q=2$ (i.e. non Gaussian corrections are negligible in this limit).
\paragraph*{}In the wave-breaking case, the stationary state is a nonequilibrium state and there is no known counterpart of\ (\ref{eq4.15}) that would make it possible to choose among the solutions to\ (\ref{eq4.14}) in a simple way. Under these circumstances, we will give all the solutions and base our discussion on the well-defined limits of a small or large $\vert\lambda\vert$ where\ (\ref{eq4.14}) has a single solution.
%
%%%%%%%%%%%%%%%%%%%%
%
\subsection{Determination of $\bm{W}$ for $\bm{\lambda <0}$ in the large system limit}\label{sec4c}
We now solve\ (\ref{eq4.14}) with $\lambda <0$ in the case of a large system for both $\varepsilon =0$ and $\varepsilon =1$. We show that the mean field theory predicts a transition from a regime of low $W$ at small $\vert\lambda\vert$ to a regime of higher $W$ at large $\vert\lambda\vert$, and that the value of $W$ at large $\vert\lambda\vert$ is significantly smaller for $\varepsilon =1$ than $\varepsilon =0$.
\paragraph*{}From now on we no longer take $L=1$. This will make $V$ appear explicitely in the self-consistency equation\ (\ref{eq4.14}). Normalizing the $a_n$ so that the right-hand side of\ (\ref{eq2.1}) is multiplied by $V^{-1/2}$, it can be seen that the right-hand side of\ (\ref{eq4.14}) must be multiplied by $V^{-1}$, and $\alpha_n$ must be replaced with $V^{1/2}\alpha_n$ on the right-hand side of\ (\ref{eq4.10}),\ (\ref{eq4.11}), and\ (\ref{eq4.14}). Note also that in the expression\ (\ref{eq4.9}) for $h_n(x)$, $k_n$ is now given by $k_n=2\pi n/L$, $n\in {\mathbb Z}^D$. In the following we will take $\alpha_n=\alpha(\vert\vert k_n\vert\vert)$ with $\alpha(k)$ the max of zero and a linearly decreasing function of $k\equiv\vert\vert {\bm k}\vert\vert$. 
\paragraph*{}We consider the large system limit in which $V$ is large enough so that discrete sums over $n\in {\mathbb Z}^D$ can be replaced by integrals over ${\bm k}\in {\mathbb R}^D$ according to
\begin{equation}\label{eq4.19}
\frac{1}{V}\sum_{n\in {\mathbb Z}^D}\rightarrow \frac{1}{(2\pi)^D}\int_{{\mathbb R}^D}d^Dk.
\end{equation}
Fix $\alpha_0>0$, $k_{max}>0$, and let
\begin{equation}\label{eq4.21}
\alpha(k)=\left\lbrace
\begin{array}{l}
\alpha_0(1-k/k_{max})\ \ {\rm if}\ k<k_{max},\\
0\ \ {\rm if}\ k\ge k_{max}.
\end{array}\right.
\end{equation}
Make the change of variables
\begin{equation}\label{eq4.20}
\begin{array}{l}
x=\mu/2 -\vert\lambda\vert W, \\
m=\mu/2,
\end{array}
\end{equation}
and define
\begin{equation}\label{eq4.22}
{\mathcal F}_{\varepsilon}(x,\beta V)=\frac{S_D}{(2\pi)^D}\int_0^{k_{max}}
\frac{1}{x+k^2}\left\lbrack 2-\frac{\beta V\alpha(k)^2}{Z_{\varepsilon}(x,k,\beta V)}\right\rbrack\, 
k^{D-1}dk,
\end{equation}
in which $S_D$ denotes the surface of the unit $D$-sphere,
\begin{equation}\label{eq4.23a}
Z_{0}(x,k,\beta V)=\frac{1}{x+k^2}\left\lbrace
\exp\lbrack\frac{1}{2}\beta V\alpha(k)^2(x+k^2)\rbrack
-1\right\rbrace ,
\end{equation}
and
\begin{equation}\label{eq4.23b}
Z_{1}(x,k,\beta V)=\left\lbrace
\begin{array}{ll}
\frac{1}{x+k^2}\left\lbrace
{\rm Ei}\lbrack\frac{1}{2}\beta V\alpha(k)^2(x+k^2)\rbrack
-\ln\lbrack\frac{1}{2}\beta V\alpha(k)^2(x+k^2)\rbrack -\gamma
\right\rbrace &{\rm for}\ \ x\ge -k^2 , \\
\frac{1}{\vert x+k^2\vert}\left\lbrace
{\rm E}_1\lbrack\frac{1}{2}\beta V\alpha(k)^2\vert x+k^2\vert\rbrack
+\ln\lbrack\frac{1}{2}\beta V\alpha(k)^2\vert x+k^2\vert\rbrack+\gamma
\right\rbrace &{\rm for}\ \ x\le -k^2 .
\end{array}
\right.
\end{equation}
Using\ (\ref{eq4.19}),\ (\ref{eq4.20}) and\ (\ref{eq4.22}), one can rewrite\ (\ref{eq4.14}) as
\begin{equation}\label{eq4.24}
m-x=\frac{q\vert\lambda\vert}{\beta}{\mathcal F}_{\varepsilon}(x,\beta V)\ \ \ \ (x\le m),
\end{equation}
the solution to which is of the form $x=x_\varepsilon(m,q\vert\lambda\vert\beta^{-1},\beta V)$. From this result,\ (\ref{eq4.20}), and the self-consistency equation $W=q\langle\vert\phi\vert^2\rangle_W$ we finally find
\begin{equation}\label{eq4.25}
\beta\langle\vert\phi\vert^2\rangle =f_\varepsilon(m,q\vert\lambda\vert\beta^{-1},\beta V)
\equiv\frac{m-x_\varepsilon(m,q\vert\lambda\vert\beta^{-1},\beta V)}{q\vert\lambda\vert\beta^{-1}}.
\end{equation}
\bigskip
\begin{figure}[htbp]
\begin{center}
\includegraphics [width=8cm] {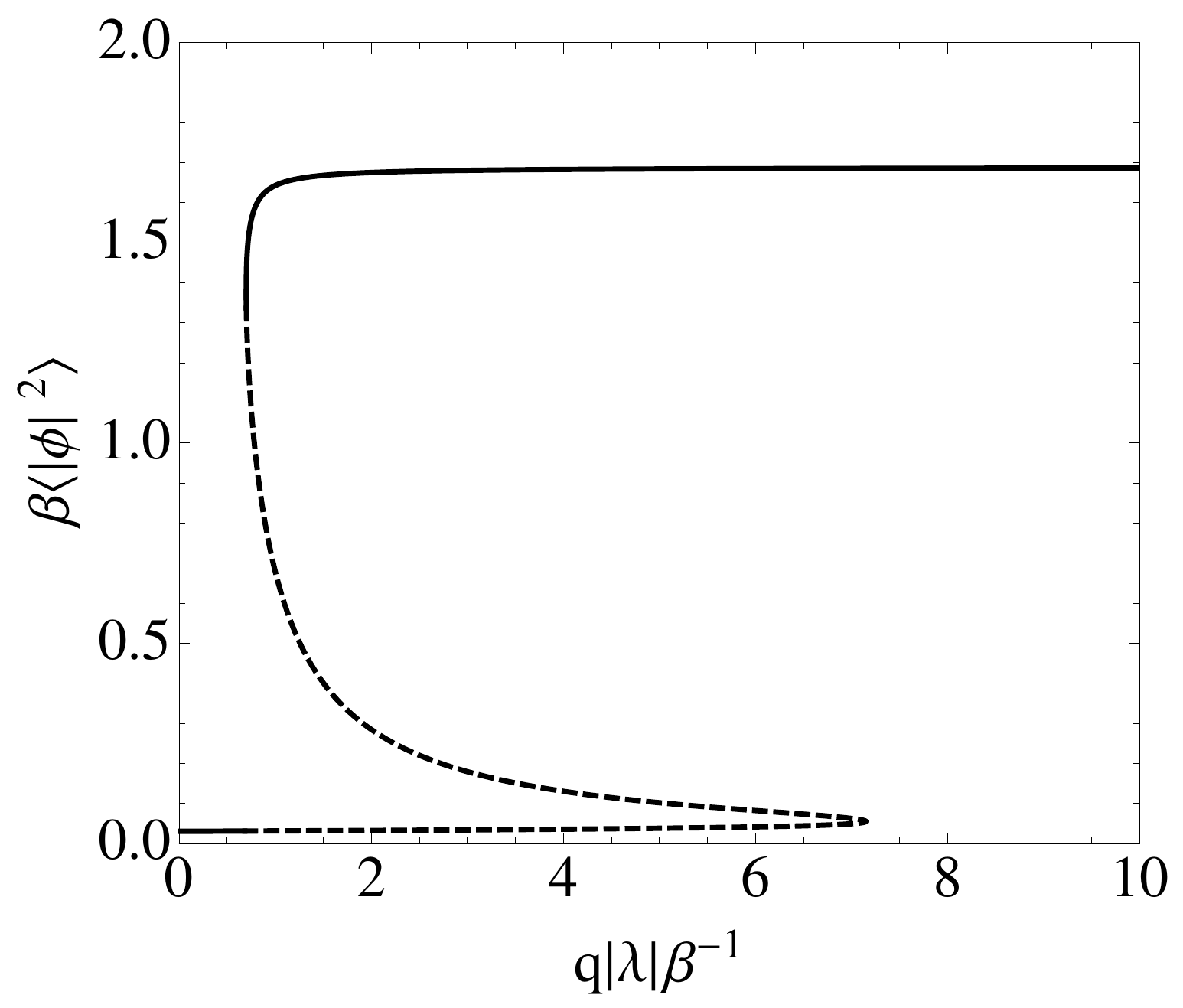}
\caption{\textsl{\small $\beta\langle\vert\phi\vert^2\rangle$ as a function of $q\vert\lambda\vert\beta^{-1}$ [Eq.\ (\ref{eq4.25})] in the reflecting boundary case ($\varepsilon =0$) for $D=3$, $\mu =1$ (i.e. $m=1/2$), $\alpha_0=1$, $k_{max}=1$, and $\beta V=10^3$. The solid line corresponds to the solution with the smallest free energy\ (\ref{eq4.16}).}}
\label{newfig1}
\end{center}
\end{figure}
\paragraph*{}Figure\ \ref{newfig1} shows a typical example of $\beta\langle\vert\phi\vert^2\rangle$ as a function of $q\vert\lambda\vert\beta^{-1}$ in the reflecting case ($\varepsilon =0$) for $D=3$, $\mu =1$ (i.e. $m=1/2$) corresponding to the value for Langmuir waves, $\alpha_0=1$, $k_{max}=1$, and $\beta V=10^3$. When\ (\ref{eq4.24}) has more than one solution, we pick the one with the smallest free energy\ (\ref{eq4.16}) (solid line). One observes a sharp transition at $q\vert\lambda\vert\beta^{-1}=0.69\pm 0.01$ from a low $\langle\vert\phi\vert^2\rangle$ at small $\vert\lambda\vert$ to a higher $\langle\vert\phi\vert^2\rangle$ at large $\vert\lambda\vert$. Similar results obtained for different values of $\beta V$, all the other parameters being fixed, show that for small $\vert\lambda\vert$ one has the usual scaling $\langle\vert\phi\vert^2\rangle\sim\beta^{-1}$. In this regime, the modes are stable and localized near $\vert a(k)\vert =0$ where they are not significantly affected by the bound on their amplitude. In the opposite case of large $\vert\lambda\vert$, the modes are unstable and localized near their maximum $\vert a(k)\vert \sim\sqrt{V}$. This is a saturated regime, independent of $\vert\lambda\vert$, and one has the simple scaling $\langle\vert\phi\vert^2\rangle\sim V$. The value of $\lambda$ at which the transition occurs scales like $\vert\lambda\vert\sim (qV)^{-1}$.
\paragraph*{}The curve for  $\beta\langle\vert\phi\vert^2\rangle$ as a function of $2\vert\lambda\vert\beta^{-1}$ for $W$ minimizing\ (\ref{eq4.16}) is indistinguishable from the solid line in Fig.\ \ref{newfig1} with $q=2$, which means that minimizing\ (\ref{eq4.16}) gives practically the same result as $W=2\langle\vert\phi\vert^2\rangle_W$. This is a consequence of the large system limit in which
\begin{eqnarray*}
\left\langle\left\vert\sum_{\vert\vert n\vert\vert <\eta_c}a_n\right\vert^4\right\rangle_W&=&
2\left\lbrack\frac{1}{(2\pi)^D}\int\langle\vert a(k)\vert^2\rangle_Wd^Dk\right\rbrack^2\\
&+&\frac{1}{V(2\pi)^D}\int\left\lbrack\langle\vert a(k)\vert^4\rangle_W-2\langle\vert a(k)\vert^2\rangle^2_W\right\rbrack\, d^Dk\\
&\simeq&2\left\lbrack\frac{1}{(2\pi)^D}\int\langle\vert a(k)\vert^2\rangle_Wd^Dk\right\rbrack^2.
\end{eqnarray*}
Using this expression in\ (\ref{eq4.16}) and minimizing w.r.t. $W$ one obtains,
$$
W\simeq\frac{2}{(2\pi)^D}\int\langle\vert a(k)\vert^2\rangle_Wd^Dk\equiv 2\langle\vert\phi\vert^2\rangle_W.
$$
\bigskip
\begin{figure}[htbp]
\begin{center}
\includegraphics [width=8cm] {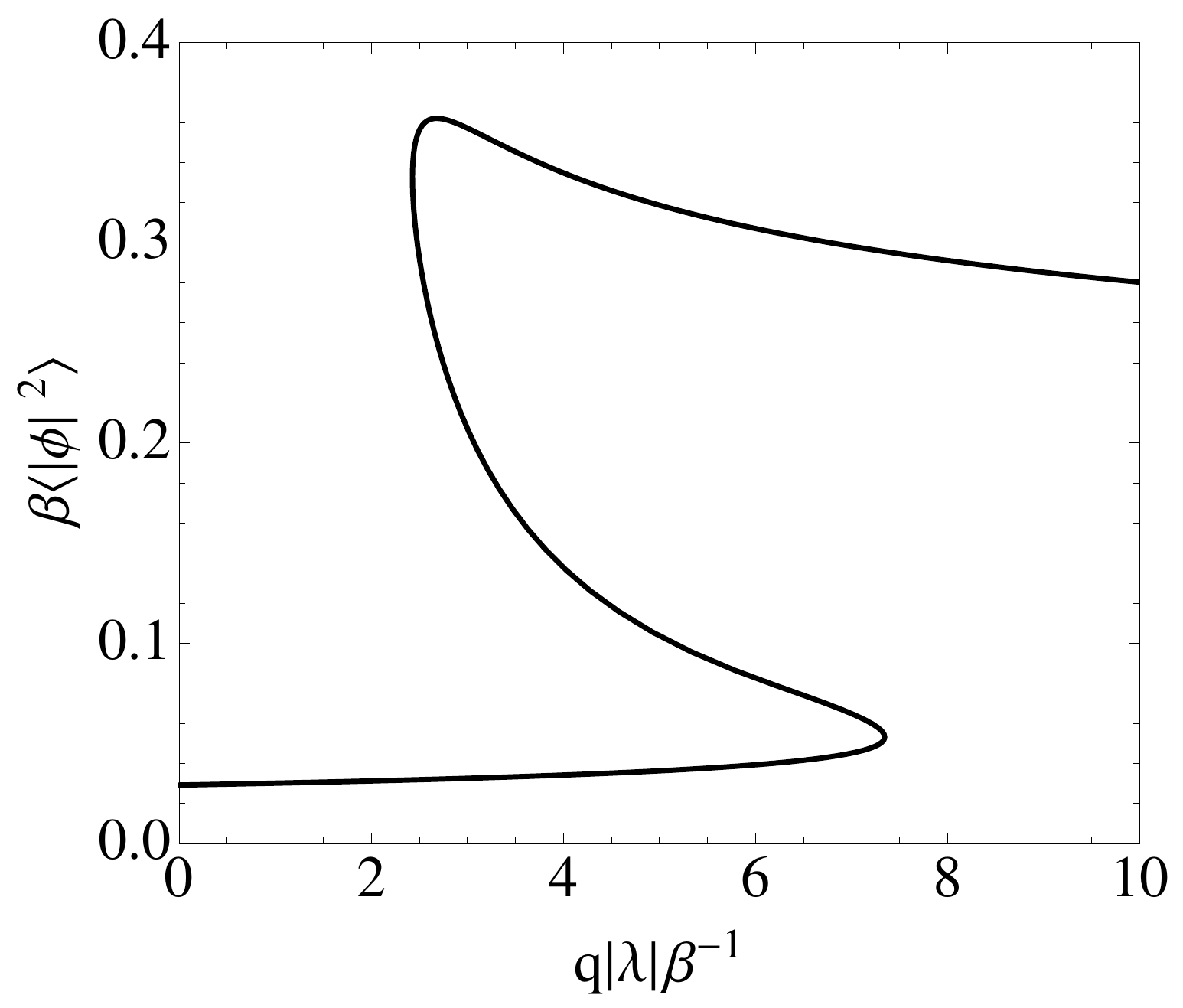}
\caption{\textsl{\small $\beta\langle\vert\phi\vert^2\rangle$ as a function of $q\vert\lambda\vert\beta^{-1}$ [Eq.\ (\ref{eq4.25})] in the wave-breaking case ($\varepsilon =1$) for $D=3$, $\mu =1$ (i.e. $m=1/2$), $\alpha_0=1$, $k_{max}=1$, and $\beta V=10^3$.}}
\label{newfig2}
\end{center}
\end{figure}
\paragraph*{}Figure\ \ref{newfig2} is the counterpart of Figure 1 for the wave-breaking case ($\varepsilon =1$) with the same parameters. The behavior of the curve at low and large $\vert\lambda\vert$, where there is only one solution, clearly suggests the existence of a transition, similar to the one observed in the reflecting case, at some intermediate $\vert\lambda\vert$. The most important difference is the strong reduction of $\langle\vert\phi\vert^2\rangle$ in the large $\vert\lambda\vert$ regime. Because of the wave-breaking prescription, unstable modes cannot remain localized near their maximum very long, as they do in the reflecting case. Instead they are rapidly reinjected at $\vert a(k)\vert =0$, which causes the observed reduction of $\langle\vert\phi\vert^2\rangle$ compared with Fig. 1. It can also be seen that $\langle\vert\phi\vert^2\rangle$ depends on $\vert\lambda\vert$ in this regime, which is thus not saturated. This is due to the fact that an unstable mode is more likely to hit its boundary, and be reset to $\vert a(k)\vert =0$, at larger $\vert\lambda\vert$. The average time it spends near its maximum is reduced, which explains the observed slow decreasing of $\langle\vert\phi\vert^2\rangle$ at large $\vert\lambda\vert$.
\paragraph*{}It must be emphasized that the phase transitions reported here are a characteristic of the mean field theory which does not exist in the exact theory for a finite $V$. The result of Section\ \ref{sec3} rules out any possibility of coexistence of distinct stationary solutions to\ (\ref{eq2.4}) at any specified $\lambda$.
%
%%%%%%%%%%%%%%%%%%%%
%
\subsection{Mean field theory with an infrared diverging $\bm{\alpha(k)}$ and $\bm{\varepsilon =1}$ in the large system limit}\label{sec4d}
For Langmuir waves in plasmas, one has both wave-breaking ($\varepsilon =1$) and an infrared diverging $\alpha(k)$ behaving like $1/k$ for $k\rightarrow 0$. Careful estimates give\ \cite{WK}\ \cite{TC}
\begin{equation}\label{eq5.1}
\alpha(k)^2\propto \frac{1+2\sqrt{3}k-(8/3^{3/4})\sqrt{k}-k^2}{k^2},
\end{equation}
for $k<1/\sqrt{3}$ and $\alpha(k)=0$ for $k\ge 1/\sqrt{3}$. The simplest way of getting meaningful results in this setting consists in getting rid of the unbounded $k=0$ mode by forcing $a_0(t)=0$ for all $t\ge 0$. For a finite $V$, all the remaining modes are bounded and the resulting mean field theory is well defined. The problem one is now faced with is the large system limit of the theory. It is easily seen from\ (\ref{eq5.1}) that $\max_{n\in {\mathbb Z}^D,\vert\vert n\vert\vert\ne 0}\alpha_n=\alpha(\min_{n\in {\mathbb Z}^D,\vert\vert n\vert\vert\ne 0}\vert\vert k_n\vert\vert)$ diverges like $V^{1/D}$ as $V\rightarrow +\infty$. This divergence jeopardizes the existence of the large system limit as it may cause an infrared divergence of $V^{-1}\sum_{\vert\vert n\vert\vert <\eta_c}\langle\vert a_n\vert^2\rangle_W$. The problem is thus to determine whether the contribution of the infrared modes to $V^{-1}\sum_{\vert\vert n\vert\vert <\eta_c}\langle\vert a_n\vert^2\rangle_W$ remains finite when the discrete sum over $n\in {\mathbb Z}^D$ is replaced by an integral over ${\bm k}\in {\mathbb R}^D$ according to\ (\ref{eq4.19}).
\paragraph*{}We consider the case in which Eq.\ (\ref{eq4.21}) is replaced with
\begin{equation}\label{eq5.2}
\alpha(k)=\left\lbrace
\begin{array}{l}
\frac{c \zeta(k)}{k}\ \ {\rm if}\ k<k_{max},\\
0\ \ {\rm if}\ k\ge k_{max}.
\end{array}\right.
\end{equation}
where $c>0$ and $k_{max}>0$ are fixed, and $\zeta(k)$ is a decreasing positive function of $k$, normalized to $\zeta(0)=1$, and such that $\zeta(k)>0$ for $k<k_{max}$ and $\zeta(k_{max})=0$.
\paragraph*{}For $x\ge 0$, $Z_{1}(x,k,\beta V)$ in\ (\ref{eq4.22}) is given by the first line of\ (\ref{eq4.23b}). In this case, the contribution of the infrared modes to\ (\ref{eq4.22}) goes like $\int_0^\xi k^{D-1}dk$ if $x>0$, and like $\int_0^\xi k^{D-3}dk$ if $x=0$, where $0<\xi\ll 1$. On the other hand, for $x<0$, the infrared contribution to\ (\ref{eq4.22}), for which $Z_{1}(x,k,\beta V)$ is now given by the second line of\ (\ref{eq4.23b}), behaves like $\int_0^\xi\vert\ln k\vert^{-1}k^{D-3}dk$.
\paragraph*{}According to these results, ${\mathcal F}_{1}(x\le 0,\beta V)$ diverges if $D\le 2$. Thus, if $\alpha(k)$ behaves like $1/k$ for small $k$, the large system limit of the mean field theory with $\varepsilon =1$ exists for $D>2$ only.
\paragraph*{}Although our proof of existence of a stationary measure for the full dynamics\ (\ref{eq2.3}) does not apply if $\alpha(k)$ is unbounded, it seems plausible that a stationary state exists for\ (\ref{eq5.1}) if $V$ is finite. Indeed, in this case the only unbounded mode is $a_0$ which is known to be modulationally unstable if its amplitude gets high enough. The larger $\vert a_0\vert$ the faster its modulational decay into bounded, smaller scale modes. Thus, it seems reasonable to expect that this transfer of energy toward smaller scales, and the subsequent dissipation according to the wave-breaking prescription, will eventually balance the growth of $\vert a_0\vert$, yielding a well defined stationary state. Proving this assertion comes within a study of the full dynamics\ (\ref{eq2.3}) with wave-breaking prescription, which is the subject of a future work.
%
%%%%%%%%%%%%%%%%%%%%
%
\section{Summary and perspectives}\label{sec5}
In this paper we have obtained stationary states for a system evolving according to a stochastic truncated NLSE on a $D$-dimensional torus, in the case of a focusing nonlinearity ($\lambda <0$). The stochastic dynamics are of a Langevin type (Ornstein-Uhlenbeck process) which act independently on every normal mode, both configurational and momentum, trying to bring its distribution to a Gibbs form at reciprocal temperature $\beta$. Our approach consists in imposing a bound on the amplitude of each Fourier mode with either reflecting or absorbing boundary conditions (with reinjection at the zero amplitude point in the latter case). The latter boundary prescription is a simple modeling of a wave-breaking-like process, like the wave-breaking of Langmuir waves in plasmas. Since there is a finite number of modes, we are left with a system with a finite number of degrees of freedom evolving according to a stochastic diffusive dynamics on a compact phase space region. It follows readily that a stationary measure always exists, is unique, and is approached exponentially fast (Sec.\ \ref{sec3}). We then studied the stationary state of a mean field version of\ (\ref{eq2.3}). We have shown that the mean field theory with $\lambda <0$ admits a transition from a regime of low field values at small $\vert\lambda\vert$ (or large $\beta$), to a regime of high field values at large $\vert\lambda\vert$ (or small $\beta$) (Sec.\ \ref{sec4}). Field values at large $\vert\lambda\vert$ are significantly smaller with wave-breaking than with reflecting boundary conditions.
\paragraph*{}The mean field theory fails to approximate the original model faithfully at the transition simply because, according to the result of Sec.\ \ref{sec3}, coexistence of distinct stationary states is impossible in a finite volume for the stochastic model of Sec.\ \ref{sec2}. This is of course similar to what happens with standard mean field theory in equilibrium where mean field predicts transitions where there is none in the real system\ \cite{LMB}. Whether or not there is a counterpart of the transition for the full dynamics of Sec.\ \ref{sec2}, as well as what the behavior of $\phi$ is like in the transition region (if it exists), are unresolved interesting questions that will probably require numerical simulations.
\paragraph*{}Our modeling of wave-breaking is an oversimplified description of the actual process. For instance, in the wave-breaking of a Langmuir wave, not all the wave energy is transferred to the particles. This could be taken into account by replacing the delta functions on the right-hand side of\ (\ref{eq2.4}) with smooth functions of small but finite width. We also make the particle degrees of freedom play the role of a reservoir for the waves. At first sight this seems reasonable since, due to the ultraviolet cut-off resulting from our wave-breaking prescription, the wave degrees of freedom are typically much fewer than the particle degrees of freedom [the ratio is of the order of $(n_e\lambda_D^D)^{-1}\ll 1$, where $n_e$ is the electron density]. Now, regarding the particles as a reservoir means that the particles should not be significantly affected by the violent exchanges of energy brought about by wave-breaking. In real situations, however, these exchanges generate a population of suprathermal electrons which can modify both the stochastic dynamics\ (\ref{eq2.3}) and the wave-breaking itself. These intermittent processes must remain rare enough so that they have a chance to always occur in a (nearly) thermalized plasma. If this is not the case, then the particles cannot be idendified with a reservoir and\ (\ref{eq2.3}) should be replaced with (i) a stochastic dynamics for the waves interacting with the particles, (ii) a stochastic dynamics for the particles interacting with the waves and a reservoir, and (iii) self-consistent wave-breaking prescriptions for (i) and (ii).
\paragraph*{}Another important simplification is the fact that the wave-breaking for a given mode ignores the existence of all the other modes (it is just a copy of wave-breaking for a single monochromatic wave). This simplification is not very realistic as particles travel in wave packets, rather than a monochromatic wave, in which several modes can have a high amplitude simultaneously. The wave-breaking of a given mode is therefore expected to be affected by the existence of the other modes {\it via} the perturbation of the electron motion they induce. What we need is a theory of wave-breaking for localized wave packets, or more generally for multimode structures. As far as we know this theory, which can be regarded as the highly nonlinear counterpart of linear Landau damping for localized wave packets, usually referred to as ``transit-time damping"\ \cite{PR1}\ \cite{PR2}, is still lacking\ \cite{note}. Having such a theory would be of the utmost interest to improve our understanding of nonequilibrium stationary states for stochastic NLSE and other stochastic nonlinear wave equations of the same type.
\section*{Acknowledgements}
We thank S. R. S. Varadhan for helpful suggestions. The work of J. L. L. and Ph. M. was supported in part by AFOSR grant 09550-07 and NSF grant DMR 08-02120.
%
%%%%%%%%%%%%%%%%%%%%
%

%
%
\end{document}